\documentclass[11pt,preprint]{aastex}
\usepackage{epsfig}
\begin{document}

\title{Hot self-similar relativistic MHD flows}

\author{Nadia L. Zakamska\altaffilmark{1,2}, Mitchell C. Begelman\altaffilmark{3,4}, Roger D. Blandford\altaffilmark{5}
\altaffiltext{1}{Institute for Advanced Study, Einstein Dr., Princeton, NJ 08540}
\altaffiltext{2}{Spitzer fellow, John N. Bahcall fellow}
\altaffiltext{3}{JILA, 440 UCB, University of Colorado at Boulder, Boulder, CO 80309-0440}
\altaffiltext{4}{Department of Astrophysical and Planetary Sciences, University of Colorado at Boulder, Boulder, CO 80309-0440}
\altaffiltext{5}{KIPAC, Stanford University, 2575 Sand Hill Road, Menlo Park, CA 94025}
}

\begin{abstract}
We consider axisymmetric relativistic jets with a toroidal magnetic field and an ultrarelativistic equation of state, with the goal of studying the lateral structure of jets whose pressure is matched to the pressure of the medium through which they propagate. We find all self-similar steady-state solutions of the relativistic MHD equations for this setup. One of the solutions is the case of a parabolic jet being accelerated by the pressure gradient as it propagates through a medium with pressure declining as $p(z)\propto z^{-2}$. As the jet material expands due to internal pressure gradients, it runs into the ambient medium resulting in a pile-up of material along the jet boundary. The magnetic field acts to produce a magnetic pinch along the axis of the jet. The relative importance of the two components -- the boundary pile-up and the magnetic pinch -- depends on the magnetization parameter of the jet and on the total energy it carries relative to the ambient pressure. Such jets can be in a lateral pressure equilibrium only if their opening angle $\theta_j$ at distance $z$ is smaller than about $1/\gamma$, where $\gamma$ is the characteristic bulk Lorentz-factor at this distance; otherwise, different parts of the jet cannot maintain causal contact. We construct maps of optically thin synchrotron emission from our self-similar models to illustrate the observational features of such jets. We suggest that the boundary pile-up may be the reason for the limb-brightening of the sub-parsec jet of M87. We find that if the synchrotron emissivity falls with the distance from the jet axis (as in the magnetically pinched case), the polarization fraction rises toward the edge, as seen in 3C273 and Mkn501. Projection effects and the emissivity pattern of the jet have a strong effect on the observed polarization signal, so the interpretation of the polarization data in terms of the geometry of magnetic fields is rather uncertain. For example, jets with toroidal magnetic fields display the `spine-sheath' polarization angle pattern seen in some BL Lac objects. 
\end{abstract}


\section{Introduction}
\label{sec_intro}

Accretion onto black holes is frequently accompanied by collimated outflows. Although many of the physical properties of these flows proved difficult to measure, strong evidence exists that they are relativistic in a variety of environments. This evidence includes the one-sided appearance of jets and observations of apparent superluminal motions in Galactic black-hole binaries \citep{mira99} and in active galactic nuclei (AGN; e.g., \citealt{jors05}), as well as the rapid gamma-ray variability in AGN \citep{dond95, ahar05}.

Jets are also magnetized. The polarization fraction of synchrotron radiation from AGN jets approaches its theoretical maximum value in some objects (e.g., \citealt{zava05, list05, push05}), indicating that the magnetic field is highly ordered. The maps of rotation measure of 3C273 \citep{asad02, zava05} and of several BL Lac objects \citep{gabu04} show strong gradients of the rotation measure across the jet which are expected if the toroidal component dominates. The polarization position angles indicate toroidal fields in more than half of all BL Lac objects \citep{gabu00, list05, push05}, but such observations may be subject to projection effects and relativistic aberration \citep{pari03, lyut05}. The strength of the magnetic field is poorly constrained, so its true dynamical importance cannot be evaluated from observations.

The current theoretical paradigm is that most relativistic jets are probably accelerated by large-scale electromagnetic stresses \citep{blan82, li92, cont94}. Acceleration is rapid until the flow passes the fast magnetosonic surface and attains rough equipartition between kinetic energy flux and Poynting flux. Further increase in the jet Lorentz-factor tends to be more gradual, but eventually it is possible for most of the Poynting flux to be converted to kinetic energy \citep{bege94, vlah03a, vlah03b, besk06, komi07}. Ultimately, the flow is confined laterally by either the inertia of the disk at the base of the jet, in which case the collimating force is transmitted along the jet by magnetic tension, or the pressure of the ambient medium along its walls \citep{bege95}. 

In this paper we investigate the interior of laminar ultrarelativistic jets whose boundary pressures are matched to their surroundings. The flux-freezing condition implies that the poloidal component of the magnetic field decays faster than the toroidal component in an expanding jet \citep{bege84}, and we consider the structure of the jet far enough from its initial acceleration site that the toroidal component dominates. \citet{cont95} showed formally that at large distances down the jet the poloidal magnetic field is dynamically unimportant and discussed the critical points of such solutions.   

As forces exterior to the jet gradually collimate it, curving the outermost streamlines slightly toward the axis, some of the material in the jet interior runs into the wall and piles up. Thus, jets may develop an edge-brightened or `hollow-cone' structure, with much of the jet energy, momentum and emissivity concentrated into a narrow sheath along the interface between the jet and its surroundings. The pinching effect due to magnetic tension has the opposite effect, tending to focus jet energy toward the axis \citep{usty95, usty00, ston00, komi07}. We find that for some functional dependencies of the ambient medium pressure $p_a(z)$ such jets can develop self-similar structures that admit a wide range of prescribed entropy and toroidal magnetic field distributions, and we investigate the interplay of pressure and magnetic field for the self-similar solutions. 

In Section \ref{sec_solutions} we describe the basic equations, our assumptions and the family of self-similar solutions that we find. In Section \ref{sec_sync} we discuss the appearance of the synchrotron radiation from our solutions. We conclude in Section \ref{sec_conclusions}. In Section \ref{sec_solutions}, unless otherwise noted we use units in which $c=1$ and the magnetic field is redefined to eliminate factors of $4 \pi$ from Maxwell's equations. The magnetic field in Gaussian units is $\sqrt{4 \pi}$ times our magnetic field ${\bf B}$ and the current (and its density) in Gaussian units is $1/\sqrt{4\pi}$ times our current (and density). Our Greek indices run from 0 to 3 and the Latin ones from 1 to 3, and summation is assumed over repeating indices (using the Minkowski metric for Greek indices). We use Gaussian and CGS units in Section \ref{sec_sync}.  

\section{Self-similar solutions}
\label{sec_solutions}

\subsection{Main equations}
\label{sec_maineq}

We start from the covariant form of the equations of relativistic MHD in flat space-time \citep{dixo78, anil89, komi99, gamm03}, assuming that there exists a frame in which the pressure tensor is isotropic:
\begin{eqnarray}
\nabla_{\nu} (\rho u^{\nu})=0; \label{covar_1}\\
\nabla_{\nu} T^{\mu\nu}=0; \label{covar_2}\\
\nabla_{\nu} (b^{\mu} u^{\nu}-b^{\nu} u^{\mu})=0.\label{covar_3}
\end{eqnarray}
These equations are, from top to bottom, the continuity equation, the energy and momentum conservation equations, and the source-free part of Maxwell's equations. Here, $\rho$ is the mass density, $u^{\mu}=(\gamma, {\bf v}\gamma)$ is the 4-velocity of a fluid element, 
\begin{equation}
T^{\mu\nu}=(\rho+\varepsilon+P+b^2)u^{\mu}u^{\nu}+(P+\frac{1}{2}b^2)g^{\mu\nu}-b^{\mu}b^{\nu} \label{en_tensor}
\end{equation}
is the energy tensor, $\varepsilon$ is the internal energy density per unit volume, $P$ is pressure, $b^2=b^{\mu}b_{\mu}$ is the electromagnetic energy density and $g^{\mu\nu}$ is the space metric (flat in our case). If ${\bf B}$ is the magnetic field (in the observer's frame) of the fluid element moving with $u^{\mu}$, the components of the 4-vector $b^{\mu}$ are $b^0=B^i u^i$ and $b^i=(B^i+b^0u^i)/u^0$, so the source-free Maxwell's equations (\ref{covar_3}) are the same as in the non-relativistic case: $\nabla \cdot {\bf B}=0$ and $\partial {\bf B}/\partial t - {\bf \nabla \times [v \times B]}=0$ (flux-freezing condition). Equations (\ref{covar_1})-(\ref{covar_3}) need to be supplemented by the equation of state of the gas $P=P(\rho,\varepsilon)$. The rest of Maxwell's equations determine the current given $b^{\mu}$ and $u^{\mu}$:
\begin{equation}
J^{\mu}=\nabla_{\nu}(e^{\mu\nu\alpha\beta}u_{\alpha}b_{\beta}),\label{eq_current}
\end{equation}
where $e^{\mu\nu\alpha\beta}$ is the Levi-Civita tensor (related to the completely antisymmetric symbol as described in \citealt{gamm03} and references therein). Equations (\ref{covar_1})-(\ref{covar_3}) are obtained under the standard assumption of ideal MHD -- that is, that the electric field in the fluid frame vanishes. The electric field in the observer's frame is non-zero, but the ideal MHD condition allows us to eliminate it from the equations (it satisfies ${\bf E}+{\bf v \times B}=0$), and we do not discuss it further. 

We now consider an ultrarelativistic gas in which the particle rest mass is small compared to the thermal energy; alternatively, our analysis is applicable to radiation-dominated flows in which the optical depth is high enough to ensure local thermodynamic equilibrium. The equation of state is then $P=\varepsilon/3$ and the mass density can be neglected in equation (\ref{en_tensor}). In this case, equations (\ref{covar_2}) and (\ref{covar_3}) do not contain the mass density which can be determined from the continuity equation given the solution $u^{\mu}$. 

We further specialize to stationary flows ($\partial/\partial t=0$). Using the condition $\nabla\cdot {\bf B}=0$ and the explicit expression $\gamma=1/\sqrt{1-{\bf v}^2}$, we rewrite the linear combination of the components of equation (\ref{covar_2}), $\gamma^4 v^i \nabla^j T^{ij}-\gamma^4 v^i v^i \nabla^j T^{0j}=0$, as 
\begin{equation}
v^i \nabla^i (P\gamma^4) + \frac{1}{2}\gamma^2 v^i \nabla^i ({\bf B}^2) - \gamma^4 v^i \nabla^j \frac{B^i B^j}{\gamma^2}=0.\label{eq_bernoulli}
\end{equation}
In the absence of a magnetic field, equation (\ref{eq_bernoulli}) implies that the quantity $S=P\gamma^4$ is conserved along any streamline and can therefore be identified with entropy, by analogy with the non-relativistic Bernoulli equation. This property is the consequence of our choice of the equation of state. In addition, in the non-magnetized case the linear combination $\gamma v^i\nabla^jT^{ij}-\gamma \nabla^j T^{0j}=0$ can be rewritten as $\nabla^i(P^{3/4}\gamma v^i)=0$, so according to equation (\ref{covar_1}) mass density in the non-magnetic flow is $\propto P^{3/4}$. 

We now consider a stationary axisymmetric ($\partial/\partial\varphi=0$) flow with a purely toroidal magnetic field ($B=B_{\varphi}$) and a purely poloidal fluid velocity. Equations (\ref{covar_2}) and (\ref{covar_3}) can be then written in cylindrical coordinates $z$ (along the axis) and $r$ (perpendicular to the axis):
\begin{eqnarray}
\frac{1}{r}\frac{\partial}{\partial r}\left[r v_r (4P\gamma^2+B^2)\right]+\frac{\partial}{\partial z}\left[v_z(4P\gamma^2+B^2)\right]=0;\label{cyl_1}\\
(4P\gamma^2+B^2)\left(v_r\frac{\partial v_z}{\partial r}+v_z\frac{\partial v_z}{\partial z}\right)+\frac{\partial}{\partial z}\left(P+\frac{B^2}{2\gamma^2}\right)=0;\label{cyl_2}\\
(4P\gamma^2+B^2)\left(v_r\frac{\partial v_r}{\partial r}+v_z\frac{\partial v_r}{\partial z}\right)+\frac{\partial}{\partial r}\left(P+\frac{B^2}{2\gamma^2}\right)+\frac{B^2}{\gamma^2 r}=0;\label{cyl_3}\\
\frac{\partial}{\partial r}(v_r B)+ \frac{\partial}{\partial z}(v_z B)=0.\label{cyl_4}
\end{eqnarray}
Equation (\ref{cyl_1}) is the 0-component of (\ref{covar_2}), whereas (\ref{cyl_2}) and (\ref{cyl_3}) are the $z$ and $r$ components of the equation $\nabla^j T^{ij}-v^i \nabla^j T^{0j}=0$ and (\ref{cyl_4}) is the $\varphi$-component of equation (\ref{covar_3}). The remaining components of equations (\ref{covar_2})-(\ref{covar_3}) are automatically satisfied given our geometric assumptions. 

From equations (\ref{cyl_1})-(\ref{cyl_4}) we see that the relativistically invariant magnetization parameter
\begin{equation}
\beta_B=\frac{B^2}{2P\gamma^2}\label{eq_betaB}
\end{equation}
(the inverse of the standard ratio of gas pressure to magnetic pressure $\beta$) quantifies the relative dynamical importance of the magnetic field. This value is also the ratio of the magnetic to thermal energy, so that $\beta_B=1$ corresponds to energy equipartition in the rest-frame of the fluid element. Indeed, for a fluid element moving with $\gamma$, the rest-frame magnetic field is $B/\gamma$ and the electric field is zero (the condition of ideal MHD), whereas pressure $P$ is a relativistic invariant. An additional factor of $4\pi$ appears in the denominator of (\ref{eq_betaB}) once the magnetic field is written in Gaussian units. In the observer's frame, the magnetic field is $B$, but there is also an electric field which counteracts the tension of the magnetic field lines, so the dynamical importance of the magnetic tension is decreased by a factor of $\gamma^2$.

We now look for self-similar solutions of equations (\ref{cyl_1})-(\ref{cyl_4}) in the paraxial ($r\ll z$) relativistic ($\gamma\gg 1$) limit, and in particular we use the approximation $v_z\simeq 1-1/(2\gamma^2)-v_r^2/2$. For streamlines, the condition of self-similarity implies that fluid elements move along the curves $r\propto z^k$. Instead of cylindrical coordinates $r$ and $z$, it becomes convenient to use the similarity variable (which is constant on a streamline) $x=r/z^k$ along with $z$. To the lowest order in $1/\gamma$, the shape of the streamline uniquely determines the radial velocity: 
\begin{equation}
v_r(x,z)=k x z^{k-1}. \label{eq_vr}
\end{equation}
We then look for all solutions of the form
\begin{equation}
\gamma(r,z)=z^{k_g}g(x), \quad P(r,z)=z^{k_p}p(x), \quad B(r,z)=z^{k_b}u(x).\label{eq_self}
\end{equation}
After transforming equations (\ref{cyl_1})-(\ref{cyl_4}) into new variables $x$ and $z$ (derivatives change as $\partial/\partial r \rightarrow z^{-k}\partial/\partial x$, $\partial/\partial z\rightarrow \partial/\partial z - (k x/z) \partial/\partial x$) we substitute the four unknown functions in the form (\ref{eq_vr})-(\ref{eq_self}) and require that $x$ and $z$ variables separate. To satisfy equations (\ref{cyl_1})-(\ref{cyl_4}), the $z$ dependence of the dominant terms should be the same. After this condition is satisfied, we write down the $x$ component of all equations. These requirements constrain combinations of $k$, $k_b$, $k_g$ and $k_p$. For example, the presence of terms $P\gamma^2+KB^2$ (where $K$ is 1/4 or 1/2) implies that $\beta_B$ (defined in eq. \ref{eq_betaB}) can depend only on $x$, so that in self-similar solutions magnetization is constant on streamlines (i.e., $k_p+2k_g=2k_b$). We then consider three cases for $v_z$: $v_r\ll 1/\gamma$, $v_r\sim 1/\gamma$ (that is, $k-1=-k_g$) and $v_r\gg 1/\gamma$. We find that there are no self-similar solutions for $v_r\gg 1/\gamma$, and we discuss this issue further in section \ref{sec_boundary}. The case $v_r\ll 1/\gamma$ yields two solutions, one with $k=0$ and one with $k=1$. The case $v_r\sim 1/\gamma$ yields another solution with $k=1/2$. In all three cases, $P\gamma^4$ is conserved along the streamlines, as in an unmagnetized case. In section \ref{sec_cyl} we discuss the solutions with straight streamlines $k=0$ (cylindrical flow) and $k=1$ (conical expansion), and in section \ref{sec_parab} the parabolic flow $k=1/2$. 

\subsection{Solutions with straight streamlines}
\label{sec_cyl}

In the case of the cylindrical jet, $k=0$, all streamlines are parallel to the axis, the radial velocity is $v_r=0$, and the corresponding self-similar solution of (\ref{cyl_1})-(\ref{cyl_4}) is $\partial P/\partial z=0$, $\partial \gamma/\partial z=0$ and $\partial B/\partial z=0$. Equations (\ref{cyl_1})-(\ref{cyl_4}) then reduce to the well-known (e.g., \citealt{komi99}) condition of balance between pressure gradients and magnetic tension forces:
\begin{equation}
\frac{{\rm d}}{{\rm d} r}P+\frac{1}{r^2}\frac{{\rm d}}{{\rm d} r}\left(\frac{r^2 B^2}{2\gamma^2}\right)=0.\label{eq_cyl}
\end{equation}
In the non-magnetic case, this condition is simply ${\rm d} P/{\rm d} r=0$, and if it is not satisfied the radial pressure gradients would create radial velocities in the flow destroying the initially cylindrical flow. The Lorentz-factor $\gamma(r)$ can vary from one streamline to another arbitrarily in the non-magnetized solution. 

In a magnetized flow given two of the functions $B(r)$, $\gamma(r)$, $P(r)$ (or two independent combinations of them) we can determine the third using equation (\ref{eq_cyl}). We assume that the magnetic field of the jet does not penetrate into the ambient medium (which we consider conductive enough). Since the total current flowing through a cross-section of radius $r$ is $I(r)=2\pi r B$, this assumption is equivalent to requiring that the net current carried by the jet is 0. The initial condition for integrating equation (\ref{eq_cyl}) is then 
\begin{equation}
\left(P+\frac{B^2}{2\gamma^2}\right)\Big |_{r_j}=P_a, \label{eq_boundary}
\end{equation}
where $P_a$ is the ambient pressure. 

\begin{figure}
\epsscale{1.0}
\plotone{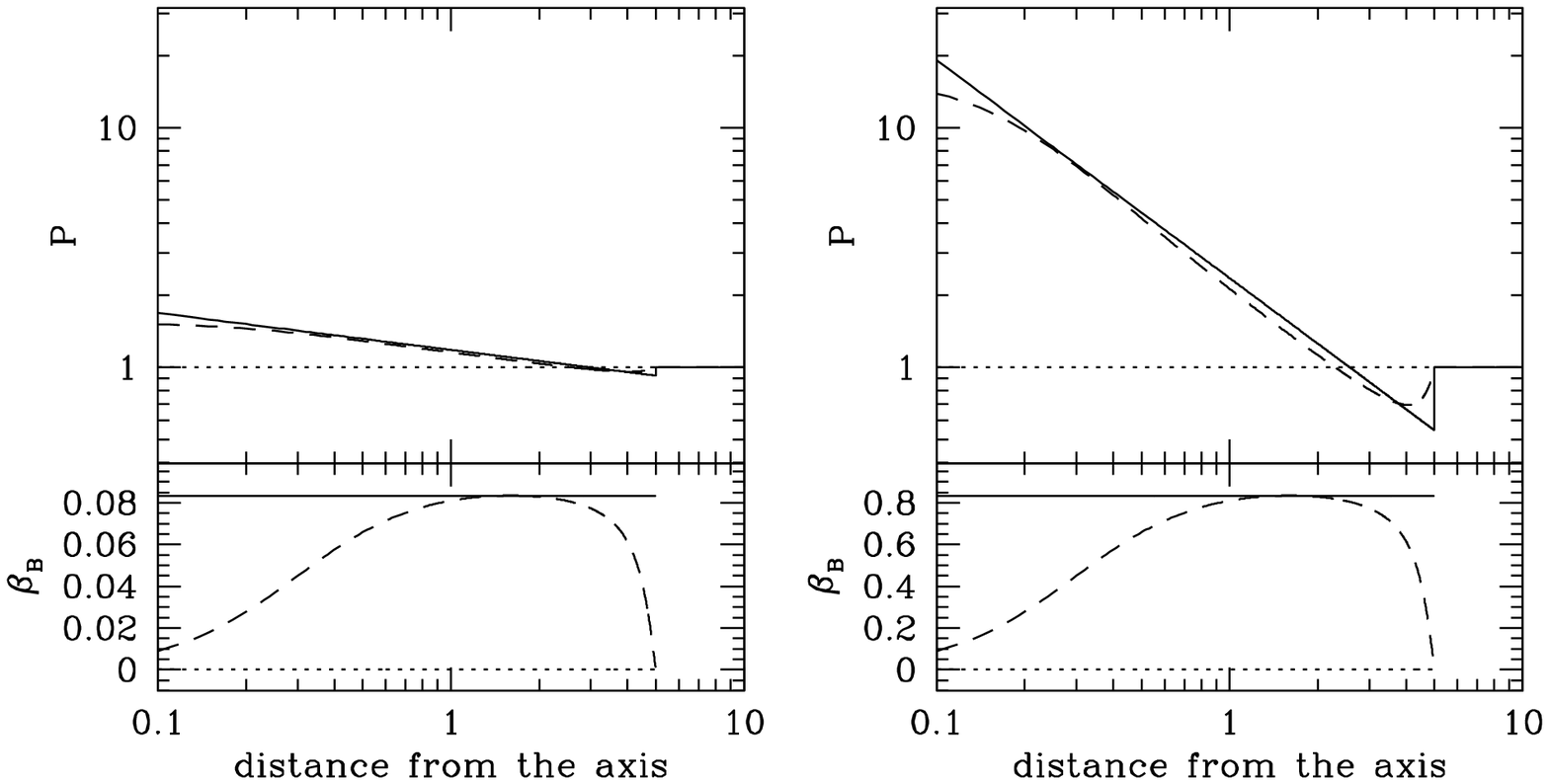}
\figcaption{Lateral structure of a cylindrical jet: non-magnetized (dotted line), constant magnetization parameter $\beta_B$ (solid line), $\beta_B(r)$ chosen so that the current density is finite everywhere (dashed line). Left panels show a weakly magnetized case ($\beta_B\la 0.1$) and right panels show a magnetized case ($\beta_B \la 1.0$). The weakly magnetized solutions differ from the magnetized solutions only by a constant stretch since for a given $\beta_B(r)$ the equation to determine pressure (\ref{eq_cyl1}) is linear. The jet boundary is at $r_j=5$.\label{pic_cyl}}
\end{figure}

As an example, we consider a flow which has a constant magnetization parameter $\beta_B(r)={\rm const}$ and is additionally homogenized so it has the same entropy on all streamlines: $S(r)=P\gamma^4={\rm const}$. In this case the solution of (\ref{eq_cyl}) is formally divergent, with infinite pressure $P(r)\propto r^{-2\beta_B/(1+\beta_B)}$ and current density $j(r)=(Br)'_r/r\propto r^{-(2+3\beta_B)/(2+2\beta_B)}$ on the axis, and the symmetry condition $B(r=0)=0$ is not satisfied. However, the divergence is weak -- the enclosed current $I(r)=2\pi r B$ and the enclosed energy flux are both zero on the axis. For a jet with a constant magnetization parameter, another singularity occurs at the jet boundary $r_j$, where a current $I_{sc}=-2\pi r_j B(r_j)$ flows with an infinite current density along the cylindrical boundary to screen the magnetic field $B(r_j)$ inside the jet. In Figure \ref{pic_cyl} we illustrate typical solutions of equation (\ref{eq_cyl}). The effect of the magnetic field is to pinch jet material toward the axis. 

The singularity on the axis arises because our equations (\ref{cyl_1})-(\ref{cyl_4}) assume that gas pressure gradient is the only force that balances the pinch due to the toroidal field. Adding other pressure components (for example, a poloidal magnetic field) would remove the singularity. Alternatively, current density and pressure singularities can be removed by assuming a finite current density everywhere and using the current $I(r)$ as a free function (along with another one, e.g., the same constant entropy condition $S(r)={\rm const}$ as above) to determine the solution of equation (\ref{eq_cyl}). Such solutions automatically satisfy the symmetry condition $B(r=0)=0$. Since there is no current singularity on the boundary of the jet in this case, the initial condition (\ref{eq_boundary}) reduces to $P(r_j)=P_a$. Solutions with finite current density are shown in Figure \ref{pic_cyl} with dashed lines. While they produce a smooth pressure profile near the axis and near the boundary of the jet, they are otherwise similar to the analytic solution. 

For the solution with $\beta_B(r)=$const, the total energy flux carried by the jet is 
\begin{equation}
\dot{E_j}=\int_0^{r_j} T^{0z} 2\pi r {\rm d} r = \int_0^{r_j} 4P\gamma^2\left(1+\frac{\beta_B}{2}\right) 2\pi r {\rm d} r = 4\pi \sqrt{SP_a(1+\beta_B)} r_j^2=4\pi P_a \gamma_j^2 r_j^2, \label{eq_enflux}
\end{equation}
which needs to be multiplied by $c$ to get the energy flux in dimensional units (cf. \citealt{lain02}). Remarkably, under our assumptions (constant entropy and magnetization across the jet) the energy flux is independent of magnetization. For AGN jets, this gives
\begin{equation}
r_j=140\mbox{ pc}\times \left(\frac{\dot{E_j}}{5\times 10^{44} \mbox{ erg sec}^{-1}}\right)^{1/2} \left(\frac{\gamma_j}{5}\right)^{-1} \left(\frac{P_a}{3\times 10^{-10}\mbox{ erg cm}^{-3}}\right)^{-1/2}.\label{eq_energy}
\end{equation}
The value for the pressure is taken from the observations of thermal X-ray emission in the center of 3C31 \citep{hard02}; it is much higher than the typical interstellar pressure in the disk of the Milky Way ($\simeq 5\times 10^{-13}$ erg cm$^{-3}$) and higher yet by comparison with intergalactic pressures. In other words, a typical intergalactic medium cannot provide enough pressure support for collimation of energetic jets. \citet{bege89} showed that such high pressures may be supplied by the cocoon which forms as the jet interacts with the ambient medium. 

Another self-similar solution of equations (\ref{cyl_1})-(\ref{cyl_4}) can be obtained for $k=1$ and corresponds to a conical flow. The streamlines are straight and the variable $x$ (which is constant on a streamline) is simply the angle this streamline makes with the axis, so that $v_r(z,x)=x$. The solution is $P(z,x)=z^{-4}p(x)$, $\gamma(z,x)=z g(x)$ and $B=z^{-1} u(x)$. The current density is also directed along the streamlines, with $j_r(z,x)=z^{-2}(xu)'_x$ and $j_z(z,x)=z^{-2}(xu)'_x/x$. The condition for maintaining the conical flow is
\begin{equation}
\frac{{\rm d}}{{\rm d} x}p+\frac{1}{x^2}\frac{{\rm d}}{{\rm d} x}\left(\frac{x^2 u^2}{2g^2}\right)=0.\label{eq_cyl1}
\end{equation}
It is identical to equation (\ref{eq_cyl}), so formally the same conclusions apply for the lateral structure of the conical jet as for the radial structure of the cylindrical jet. However, in this geometry the jet is expanding fast enough that the outer parts of it are causally disconnected from the inner parts. We discuss the applicability of our solutions to this case in Section \ref{sec_boundary}.

\subsection{Parabolic solution}
\label{sec_parab}

\begin{figure}
\epsscale{0.9}
\plotone{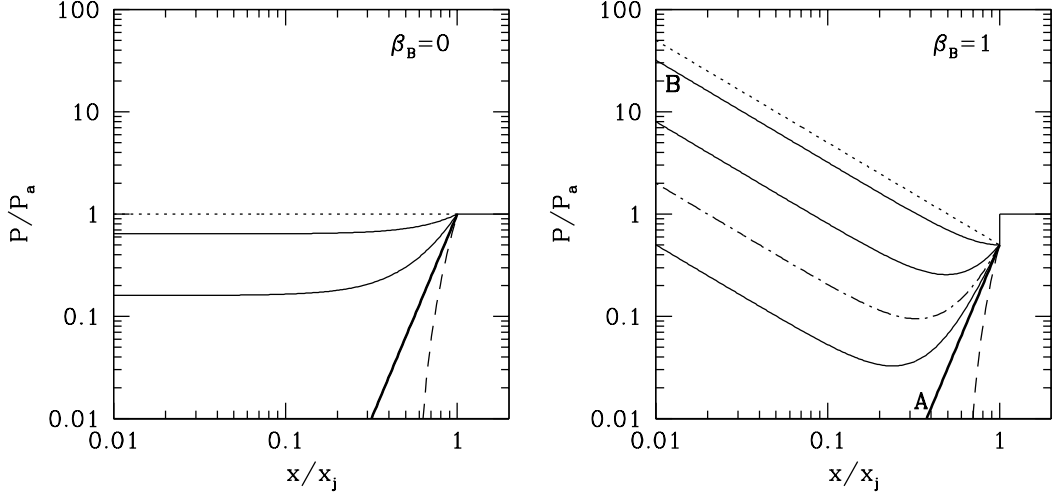}
\figcaption{Lateral structure of a parabolic jet: non-magnetized (left) and magnetized with a constant magnetization parameter $\beta_B=1$ (right). In each panel, the different lines, from top to bottom, are analytical solutions of equation (\ref{eq_par}) with increasing value of the Lorentz-factor (parametrized by $g_j$), with all other parameters ($\beta_B$, $p_a$ and $x_j$) fixed. The dotted lines show the cylindrical solution for comparison; the solid and dot-dashed lines show physically acceptable solutions $g_j<g_j({\rm cr})$, where the critical value is given by (\ref{eq_unconnected}). The dashed lines show an unphysical solution $g_j>g_j({\rm cr})$ for which pressure formally becomes 0 near the axis. The thick solid lines show the critical solution with $g_j=g_j({\rm cr})$. In the non-magnetized case (left) the energy carried by the jet monotonically increases downward, whereas in the magnetized case (right) it reaches its maximum value given by equation (\ref{eq_emax}) on the dot-dashed curve and then declines. Labels A and B indicate the models used in calculation of synchrotron emission in Figure \ref{pic_map}. \label{pic_parab}}
\end{figure}

There is only one more solution of equations (\ref{cyl_1})-(\ref{cyl_4}) in the form (\ref{eq_vr})-(\ref{eq_self}), and it occurs at $k=1/2$, so that the streamlines are parabolae ($r=xz^{1/2}$). In this case, the functions vary along the streamlines as $v_r(z,x)=z^{-1/2}x/2$, $P(z,x)=z^{-2}p(x)$, $\gamma(z,x)=z^{1/2}g(x)$, $B(z,x)=z^{-1/2}u(x)$. The lateral structure of the jet satisfies
\begin{equation}
\frac{{\rm d}}{{\rm d} x}p+\frac{1}{x^2}\frac{{\rm d}}{{\rm d} x}\left(\frac{x^2 u^2}{2g^2}\right)-x\left(pg^2+\frac{u^2}{4}\right)=0.\label{eq_par}
\end{equation}
Just like in the cylindrical and conical cases, the first two terms describe the balance between pressure gradients and magnetic tension forces, whereas the last term represents centrifugal forces which arise because the fluid moves along curved streamlines in the parabolic case. Again, there are two free functions and the third can be determined from equation (\ref{eq_par}). In the case when the free functions are $s(x)=p g^4={\rm const}$ and $\beta_B(x)=u^2/2pg^2={\rm const}$, the solutions can be obtained analytically, with the initial condition (\ref{eq_boundary}) at the boundary of the jet $x_j=r_j/z^{1/2}$. Current density singularities similar to those discussed in the previous section are also present in this case, but since they are weak (in the sense that the total current and energy flowing through a finite cross-section are finite) and can be easily eliminated by an appropriate choice of the current density, we do not concern ourselves with them any further. The external pressure must decline as $P_a(z)=p_a z^{-2}$ as the distance $z$ from the source increases in order for the parabolic solution to be realized. Example solutions for the lateral structure are shown in Figure \ref{pic_parab}. 

In the non-magnetized case (left panel in Figure \ref{pic_parab}), the first free function is $u=0$, and we again consider an isentropic flow $s(x)=p g^4={\rm const}$. The solution of equation (\ref{eq_par}) is then 
\begin{eqnarray}
\sqrt{\frac{p(x)}{p_a}}=1-\frac{g_j^2 x_j^2}{4}\left(1-\frac{x^2}{x_j^2}\right). \label{eq_pressure}
\end{eqnarray}    
These solutions are shown in Figure \ref{pic_parab} with solid lines. However, for $g_j>2/x_j$ (or alternatively $s>s_{\rm cr}=16 p_a/x_j^4$) the right-hand side of this equation becomes negative near the axis, and we conclude that for these values of the Lorentz-factor (entropy) a global self-similar solution of MHD equations (\ref{cyl_1})-(\ref{cyl_4}) does not exist. The reason for this behavior is clear when we rewrite this condition as $\gamma_j(z)>2 z/r_j(z)$. In other words, the global solution in the parabolic case does not exist if $\gamma$ is larger than, roughly, the inverse of the opening angle of the jet $1/\theta_j$. In this case the outer parts of the jet are not causally connected with the inner parts, and the lack of a global solution is unsurprising. 

In the magnetized case (right panel in Figure \ref{pic_parab}), the main difference from the purely hydrodynamic case is the pinching effect of the magnetic field. The constant entropy and magnetization solutions are 
\begin{equation}
\sqrt{\frac{p(x)}{p_a/(1+\beta_B)}}=\left(\frac{x}{x_j}\right)^{-\beta_B/(1+\beta_B)}\left(1-\frac{g_j^2 x_j^2}{4}\frac{1+\beta_B/2}{1+3\beta_B/2}\right)+\frac{g_j^2 x^2}{4}\frac{1+\beta_B/2}{1+3\beta_B/2},\label{eq_pressure_mhd}
\end{equation}
which turn into solution (\ref{eq_pressure}) for $\beta_B=0$. The entropy of this solution is $s=p_ag_j^4/(1+\beta_B)$. The pinching effect does not eliminate the same critical behavior as we see in the non-magnetized case, and for
\begin{equation}
g_j>g_j({\rm cr})=\frac{2}{x_j}\sqrt{\frac{1+3\beta_B/2}{1+\beta_B/2}} \label{eq_unconnected}
\end{equation}
the global solution does not exist. 

The total energy carried by the jet can be calculated the same way as in equation (\ref{eq_enflux}):
\begin{equation} 
\dot{E}_j=4\pi p_a g_j^2 x_j^2-\frac{\pi}{2}p_a\frac{1+\beta_B/2}{1+\beta_B}g_j^4 x_j^4\label{eq_enflux_parab}
\end{equation}
(to be multiplied by $c$ to get the energy flux in CGS units; $p_a$, $g_j$ and $x_j$ are all dimensional values). We now fix the values of $\beta_B$, $p_a$ and $x_j$ and consider the solutions for different values of $g_j$. The behavior of the solutions is illustrated in Figure \ref{pic_parab}, with $g_j$ increasing from the top curve to the bottom curve in both panels. In the hydrodynamic case ($\beta_B=0$; Figure \ref{pic_parab}, left) as we increase $g_j$ the total energy of the jet $\dot{E}_j$ increases all the way up to the point described by condition (\ref{eq_unconnected}), when the jet becomes too fast to maintain causal contact and solutions (\ref{eq_pressure_mhd}) become inapplicable. For the magnetized jet ($\beta_B>0$; Figure \ref{pic_parab}, right), as we consider solutions with increasing values of $g_j$, the total energy of the jet also increases at first, but then reaches its maximum value of
\begin{equation}
\dot{E}_{\rm max}=8\pi p_a\frac{1+\beta_B}{1+\beta_B/2}\label{eq_emax}
\end{equation}
(times $c$ in CGS units) and then decreases slightly before the jet becomes causally disconnected. For energies close to $\dot{E}_{\rm max}$, most of the jet material is pushed against the wall of the jet, whereas for small energies most of the jet material is in the pinched core. One can think of increasing the Lorentz-factor of the flow as analogous to increasing the flow's inertia, so that the greater the inertia, the harder it is to pinch the jet toward the axis. 

\subsection{Comments on flow geometry and causality}
\label{sec_boundary}

While we do see a pinching effect of the magnetic field in the self-similar parabolic solution, we need to make a distinction between our solution and the electromagnetic collimation seen in 2D numerical simulations. In the numerical simulations by \citet{komi07} and others the core is almost cylindrical when the outer streamlines are parabolic, so the ratio $\theta_{\rm core}/\theta_j$ decreases away from the base. The magnetic fields act to provide an additional collimation of the jet matter (the poloidal component of the magnetic field is dynamically important), not just to redistribute the matter close to the axis. In our case all streamlines are parabolic, and the pinched core also has a parabolic shape (albeit with a smaller opening angle than the outer streamlines), so that $\theta_{\rm core}/\theta_j=$const. The difference between the geometry of our solutions and those obtained in numerical simulations is a direct consequence -- and the limitation -- of the self-similarity assumption, in that no additional collimation beyond what is provided by the ambient pressure can be obtained in the self-similar solutions. In our solution the $z$-dependence of all functions, including $\gamma(x,z)=g(x)z^{1/2}$, is identical in the magnetized and purely hydrodynamic cases, indicating that acceleration by electromagnetic forces is matched to the thermal acceleration (their ratio is a constant on a streamline). 

The lack of causal connection for $\theta_j\gamma>1$, which we suggested in Section \ref{sec_parab} based on the lack of a global solution of the lateral equilibrium equation, can be understood if one considers the expansion of the jet cross-section in the frame moving with the Lorentz-factor $\gamma$. In this frame, the condition $\theta_j\gamma<1$ is the same as the requirement that the cross-section expands slower than the speed of light. In the observer's frame, the condition $\theta_j\gamma<1$ comes from requiring that the time-scale for signal propagation across jet at distance $z$ is smaller than the time scale for the fluid elements to get from $z$ to $2z$. 

Whether the condition $\theta_j\gamma<1$ is satisfied depends not just on the physical parameters such as energy or magnetization, but also on the general geometry. For the cylindrical solution, the value $\theta_j\gamma$ decreases with the distance from the origin of the jet, since $\gamma=$const. Therefore, far enough from the origin the cylindrical jet is in causal contact. For the conical jet, $\gamma\propto z$ and $\theta_j=$const, so far enough from the origin the conical jet always loses causal contact. When that happens, the condition $v_r\ll 1/\gamma$ used to derive the conical solution (end of section \ref{sec_maineq}) fails, since $v_r=\theta_j$. Therefore, the self-similar conical flow whose lateral structure satisfies equation (\ref{eq_cyl1}) is no longer a solution of the MHD equations downstream of $\theta_j\gamma\sim 1$. The parabolic case is unique in that the product $\theta_j\gamma$ remains constant as the jet propagates. In other words, in the parabolic case the sound-crossing time of the jet cross-section is matched to its expansion and propagation rate. 

Our critical condition (\ref{eq_unconnected}) was obtained for a specific choice of free functions, namely, $s(x)=$const and $\beta_B(x)=$const. We now consider how this choice affects the presence of a critical condition and we take as an example a solution with a different set of free functions: $g(x)=$const and $\beta_B(x)=$const. The solution is then
\begin{equation}
p(x)=\frac{p_a}{1+\beta_B}\left(\frac{x}{x_j}\right)^{-2\beta_B/(1+\beta_B)}\exp\left(\frac{1+\beta_B/2}{2(1+\beta_B)}(x^2-x_j^2)g^2\right).
\end{equation}
The qualitative features of this solution are the same as those of solution (\ref{eq_pressure_mhd}), as both the edge pile-up (exponential term) and the magnetic pinch (power-law term) are present. Although the global solution formally exists for all $x_j$ and $g$, for large values of $x_j g$ all of the matter is piled up into the layer of thickness $\Delta x \sim 1/g$ near the edge and the pressure is exponentially small elsewhere. 

In this paper we investigate the structure of the jet for three special cases of the ambient pressure ($p\propto z^0$, $p\propto z^{-2}$ and $p\propto z^{-4}$). In general, the interactions between the jet and the ambient medium may result in a formation of a shocked layer, such as that described by \citet{brom07} for a hydrodynamic stationary jet. Self-similar jets should be considered special cases in which a solution without a shocked layer is possible. 

\section{Synchrotron emission}
\label{sec_sync}

\subsection{Constructing emission and polarization maps}

Our models relate the distribution of pressure and the magnetic field within the jet. In this setup, the relativistic particles that produce the pressure emit synchrotron radiation. If the total synchrotron losses are negligible relative to the energy flux of the jet, this emission does not affect the dynamical structure of the jet, and we can calculate the spatial distribution of synchrotron emission given $P(r,z)$ and $B(r,z)$. 

We assume that all emission is optically thin and that the pressure in our model is due to relativistic particles with the distribution function $N(E){\rm d}E=K_e E^{-2}{\rm d}E$ between $E_{\rm min}$ and $E_{\rm max}$, so the spectral index of the synchrotron emission is $\alpha=0.5$ (defined as $S_{\nu}\propto\nu^{-\alpha}$) and the pressure is $P=\varepsilon/3=K_e\ln(E_{\rm max}/E_{\rm min})/3$. Since the pressure is rather insensitive to the poorly known value $E_{\rm max}/E_{\rm min}$, the normalization of the distribution function $K_e$ follows the pressure during the adiabatic expansion as the jet propagates. Assuming a purely leptonic jet, the rest-frame emissivity per unit frequency and unit solid angle is 
\begin{equation}
j'_{\nu'}(\nu', {\bf n'})=4.86\times 10^{-14} |{\bf b \times n'}|^{3/2} K_e \nu'^{-0.5}\mbox{ erg s}^{-1}\mbox{ ster}^{-1}\mbox{ cm}^{-3}\mbox{ Hz}^{-1},\label{eq_emis}
\end{equation} 
where $b$ is the rest-frame magnetic field in Gauss, $K_e$ is in erg cm$^{-3}$ and rest-frame frequency $\nu'$ is in Hz (e.g., \citealt{ginz79}). The unit vector in the direction of propagation of a photon is ${\bf n}$ in the observer's frame, but the same photon propagates with ${\bf n'}$ in the rest-frame of the fluid element. The rest-frame magnetic field is $b=B/\gamma=\sqrt{8\pi\beta_B P}$, where we are now using the Gaussian units. For a constant magnetization parameter, the emissivity would have traced pressure as $j_{\nu}\propto P^{(3+\alpha)/2}=P^{7/4}$ (as in the classical equipartition case, \citealt{bege84}) if not for the relativistic beaming and aberration effects, which are different for every volume element since they all move with different velocities at different angles to the line of sight. 

We integrate emissivity (\ref{eq_emis}) along the line of sight (parametrized by $\zeta$) in the observer's frame and at a fixed observer's frequency $\nu$ (corresponding to different frequencies $\nu'$ in the rest-frame). The observed flux density from a piece of jet with coordinates $\xi$, $\eta$ as projected on the plane of the sky subtending a solid angle of $\Delta\Omega$ is
\begin{equation}
S_{\nu}(\xi, \eta)=\Delta\Omega \int \delta^{2+\alpha}(\xi,\eta,\zeta)j'_{\nu'}(\nu, \xi,\eta,\zeta, {\bf n'}) {\rm d}\zeta,\label{eq_flux}
\end{equation}
where the Doppler factor is $\delta=[\gamma(1-{\bf v \cdot n}/c)]^{-1}$. The aberration factor (the angle $\psi'$ between ${\bf b}$ and ${\bf n'}$ in eq. \ref{eq_emis}) can be calculated using the Lorentz-transformation of the propagation vector (Appendix C of \citealt{lyut03}) and turns out to be $\sin^2 \psi'=1-\delta^2\cos^2\psi$.

In addition to the intensity map (\ref{eq_flux}), we construct maps of polarization position angle and polarization fraction. To this end, we calculate the emissivity into the two independent modes of linear polarization, that is, Stokes parameters $j_Q$ and $j_U$ emitted by a unit volume within the jet \citep{ginz79, lyut03}. These expressions are similar to (\ref{eq_emis}), but contain additional dependence on the orientation of the magnetic field through the $E$-vector of polarization. As detailed in \citet{lyut03}, this latter vector also suffers relativistic aberration which we take into account in our calculations. The maps of $Q(\xi,\eta)$ and $U(\xi,\eta)$ are then obtained just like the total intensity (\ref{eq_flux}). Polarization fraction is $\sqrt{Q(\xi,\eta)^2+U(\xi,\eta)^2}/S_{\nu}(\xi,\eta)$ and polarization position angle is $\arctan [U(\xi,\eta),Q(\xi,\eta)]/2$. 

Five parameters appear in equations (\ref{eq_pressure_mhd}), (\ref{eq_emis}) and (\ref{eq_flux}): $\phi$ is the angle between the jet axis and the line of sight; $p_a$ is the normalization of the ambient pressure; $\beta_B$ is the magnetization parameter; $x_j$ is the normalization of the opening angle; and $g_j$ is the normalization of the bulk Lorentz-factor. Solution (\ref{eq_pressure_mhd}) has a linear dependence on $p_a$, so changing this parameter simply results in scaling the total intensity $\propto p_a^{7/4}$. Furthermore, because of the self-similarity of our solutions $x_j$ is a geometrical factor, and increasing $x_j$ results in zooming in on the region near the basis on the jet. Therefore, if $g_j$ is expressed as a fraction of its critical value given by (\ref{eq_unconnected}) then $p(x/x_j)$ (eq. \ref{eq_pressure_mhd}) does not depend on $x_j$. As a result, only $\phi$, $\beta_B$ and $g_j/g_j({\rm cr})$ need to be specified to produce maps of synchrotron emission, and  $p_a$ is necessary for determining the overall flux normalization.

\subsection{Edge-brightened jet}

\begin{figure}
\epsscale{0.9}
\plotone{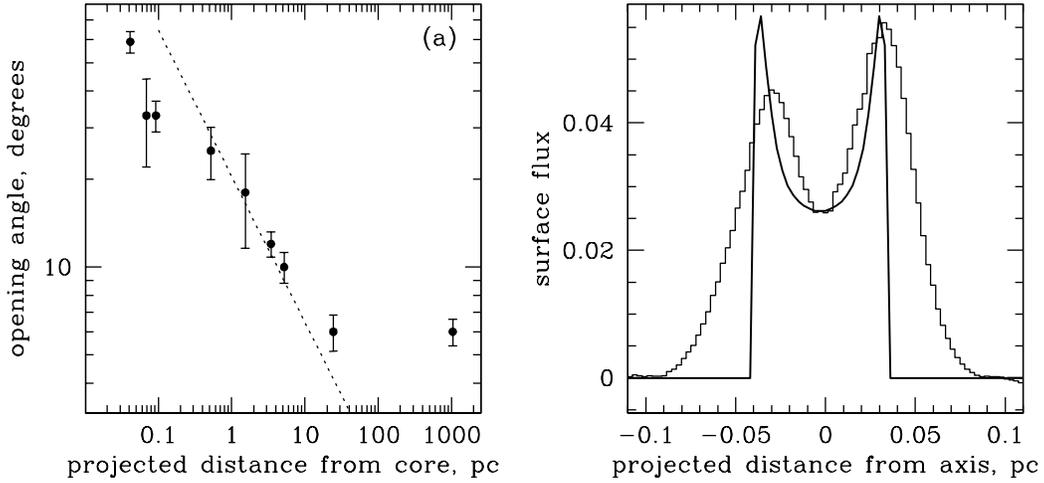}
\figcaption{(a) Opening angle of the jet in M87 as a function of distance from the core. The points with error bars are from \citet{juno99} and references therein. For comparison, the dotted line shows the opening angle of a parabolic jet ($\theta_j\propto 1/\sqrt{z}$); it is not a fit to the data. (b) Lateral surface brightness profile of the jet in M87. The profile constructed from data by \citet{ly07} is shown as a histogram. The profile from our model is shown with a thick solid line. Effects of the finite beam size have not been taken into account. \label{pic_m87} }
\end{figure}

As an example, we consider the jet in the nucleus of a nearby elliptical galaxy M87. The large mass of the supermassive black hole which powers the jet ($\sim 3\times 10^9 M_{\odot}$) and its relatively small distance (16 Mpc) imply that the Schwarzschild radius $r_s$ subtends a few micro-arcsec on the sky, bringing the immediate environment of the black hole within reach of modern radio observations. The VLBI images on scales of few hundred $r_s$ show that the jet is wide at the base and collimated at large distances from the nucleus \citep{juno99, ly07, kova07}; in fact, the jet is rather close to being parabolic in shape (Figure \ref{pic_m87}a). Additionally, it is strongly edge-brightened, which is usually interpreted in terms of the Kelvin-Helmholtz instability acting on the boundary between the jet and the surrounding medium (\citealt{aloy00, peru07}; see also discussion of various models in \citealt{kova07}). We propose that the edge pile-up seen in our models of pressure-matched jets may provide an alternative explanation for edge-brightening in the sub-pc jet of M87. 

Among our models, the most strongly edge-brightened one corresponds to the critical solution $g_j=g_j({\rm cr})$ (eq. \ref{eq_unconnected}; model A in Figure \ref{pic_parab}) which we adopt as a model for the jet in M87. In this case the magnetization parameter $\beta_B$ enters the expression for $g_j({\rm cr})$ and therefore weakly affects the Doppler factor, but the pressure profile (\ref{eq_pressure_mhd}) does not explicitly depend on $\beta_B$. Therefore, the spatial distribution of emission is almost independent of $\beta_B$, while the total intensity scales as $\propto\beta_B^{3/4}$ (eq. \ref{eq_emis}). We take $\beta_B=1$. The last parameter needed to specify the model is the angle between the jet axis and the line of sight $\phi$. In M87 it is constrained to be between $30^{\rm o}$ and $50^{\rm o}$ by the combination of apparent velocities of clumps (assumed to be representative of bulk fluid motions) and the brightness ratio of the approaching main jet to the receding counterjet \citep{ly07}. If we further assume that the jet is relativistic ($v/c\simeq1$) by the time it reaches the projected distance of 3.1 mas and if we take the brightness ratio measurement of \citet{kova07} at face value ($R>200$ at this distance) we obtain $\phi<38^{\rm o}$. We use $\phi=30^{\rm o}$ in our models. The observed emission from the jet of M87 is optically thin ($\alpha=0.4\pm 0.3$; \citealt{ly07}), in agreement with the assumptions we made when constructing emissivity maps. 

\begin{figure}
\epsscale{0.85}
\plotone{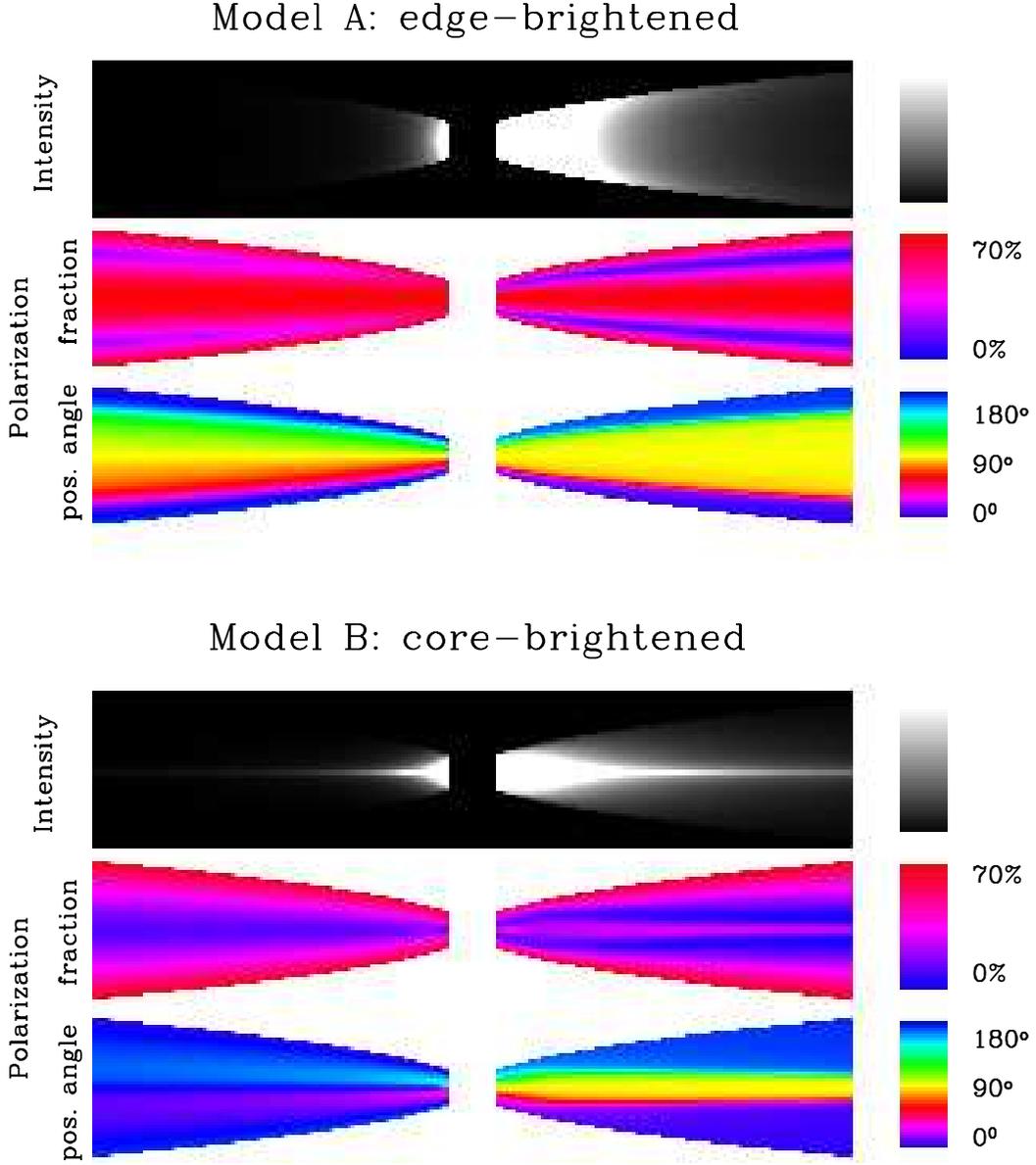}
\figcaption{Maps of synchrotron emission from our models. Model A is for the critical solution (thick solid line in Fig. \ref{pic_parab}b), in which all of the material is piled up against the wall of the jet, whereas model B is for the core-dominated solution (top solid line in Fig. \ref{pic_parab}b). For each of the two models, we show intensity (square root scaling), polarization fraction (blue -- low, red -- high) and polarization position angle counted clockwise from the vertical direction (yellow -- polarization is along the axis, blue -- polarization is perpendicular to the axis). The core region is excluded from the maps. The right side of the jet is moving toward the observer. The Lorentz-factor of the fluid follows $\gamma\propto P^{-1/4}$; the normalization is such that for model A, $\gamma \simeq 15$ for the fluid elements in the top right corner of the intensity map, and for model B, $\gamma \simeq 3$ for the same elements. \label{pic_map}}
\end{figure}

The maps of total intensity, polarization fraction and polarization position angle for this setup are shown in Figure \ref{pic_map}A. The brighter right side is the component of the jet moving toward the observer. The contrast in intensity between the edge and the emission observed on-axis is similar in our model and in the jet of M87 (Figure \ref{pic_m87}b). The brightness of the jet in the intensity map declines away from the base of the jet roughly as $\propto \xi^{-3.8}$, where $\xi$ is the coordinate along the projected axis. This decline is due to the combined effects of the adiabatic expansion and the changes in aberration and the Doppler factor as the jet accelerates. The observed emission declines somewhat slower, $\propto \xi^{-2.4}$ as measured between the clumps from the map provided by \citet{ly07}. One possible explanation for the difference in decline rates, if they are taken at face value, is that our models are developed in the ultrarelativistic approximation and are not strictly applicable to the jet in M87 which has a typical Lorentz-factor of just a few, especially near the base. Comparing the model intensity to the observed intensity at about 2 mas from the base of the jet (projected distance of 0.16 pc$=500 r_s$) leads to an estimate of the ambient pressure ($\sim 10^{-3}$ erg cm$^{-3}$), of the total energy flux carried by the jet (a few times $10^{44}$ erg sec$^{-1}$; eq. \ref{eq_enflux_parab}) and of the magnetic field ($\sim 1$ Gauss). 

Such estimates are based on the standard assumption that the observed synchrotron emission from the jet reflects bulk motions and bulk pressures of the material in the jet, but it is not clear that this is indeed the case. An extreme example is the radio jet in Centaurus A, in which practically all of the jet emission is contained in compact clumps \citep{hori06}. Although the individual exposures of the jet in M87 are clumpier than the time-averaged composite that we used for our analysis \citep{ly07}, they all show the same edge-brightened structure in the process of collimation, lending some support to the bulk origin of this emission. Because the nature of clumps and their relationship to the bulk motions of the fluid are not understood, in our model we do not attempt to reproduce the somewhat contradictory data on proper motions of these features \citep{ly07, kova07}. 

\begin{figure}
\epsscale{0.8}
\plotone{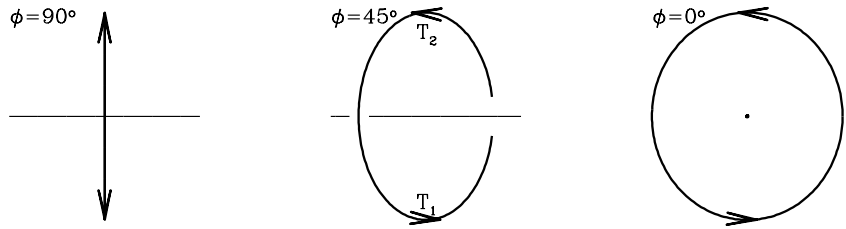}
\figcaption{Appearance of a toroidal field line (thick) and the axis of the jet (thin) as projected on the sky for a set of inclination angles $\phi$. Even though the field is purely toroidal, for all angles $\phi\ne 90^{\rm o}$ it will appear as parallel to the axis in the tangential points $T_1$ and $T_2$. \label{pic_loop}}
\end{figure}

Polarization measurements for the jet in M87 are now being obtained \citep{ly07}, and we present polarization maps expected from our models in Figure \ref{pic_map}A. The polarization fraction of emission from every volume of fluid is the same and equals $(1+\alpha)/(\frac{5}{3}+\alpha)\simeq 0.69$, but the polarization position angles of emission from different elements of the volume vary, so the polarization signal washes out when magnetic fields with different directions are present along the same line of sight. The polarization fraction and the polarization position angle are not the same in the jet and in the counterjet because of the effects of aberration mentioned above, even though the purely geometrical projection effects are the same for the main jet and for the counterjet. In the edge-brightened case the polarization is expected to be high almost everywhere. The orientation of the magnetic field changes along every line of sight, but only a thin region near the edge contributes most of the emission, and the change in the direction of ${\bf B}$ across this layer is small. Stripes of low polarization seen in the map occur because the superposition of emission from the near and far walls of the jet leads to depolarization in these areas. On the projected axis of the jet, the magnetic field in the emitting region lies in the plane of the sky and is perpendicular to the axis, so the polarization $E$-vector is parallel to the axis. On the edge of the jet, the projected magnetic field is parallel to the axis (Fig. \ref{pic_loop}), so the polarization $E$-vector is perpendicular to the axis. In other words, even though the magnetic field is purely toroidal, the projection effects result in the so-called `spine-sheath' polarization angle pattern \citep{attr99,push05}. This pattern is sometimes interpreted as evidence for a poloidal component of the magnetic field near the edge of the jet (for example, due to shearing between the jet and ambient medium). But as long as the jet does not lie exactly in the plane of the sky, projection effects lead to the `spine-sheath' pattern even for a purely toroidal field.

\subsection{Core-dominated jet}

As another example, we now consider the other extreme in which most of the emission is coming from the pinched core. In this case the surface brightness on the axis formally diverges for $\beta_B>2/5$. One could eliminate this singularity by considering different profiles of $\beta_B(x)$, such as those that we describe in Section \ref{sec_cyl}. Alternatively, it can be shown that the optical depth for synchrotron self-absorption (opacity is $\propto K_e \nu'^{-3} b^2$) diverges faster than emissivity for these values of $\beta_B$. So the emission from the pinched core is composed of a geometrically thin optically thick core surrounded by an optically thin jet. If we look at the jet at higher frequencies, we see deeper into the jet, so the emissivity near the optically thick core rises steeply, but at the same time the surface area of the optically thick core shrinks. The result is that one can always find a frequency at which the optically thick core contributes a negligible amount to the total emission. Observed jets are typically optically thin, so we assume that this condition is satisfied. The value of surface brightness exactly on axis is affected by the effects of optical depth, so in our calculations we assume that the core is smaller than a resolution element of our map and do not consider the lines of sight going directly through the axis. 

The resulting maps of total intensity, polarization fraction and polarization position angle are presented in Figure \ref{pic_map}B. As expected, the intensity rises toward the axis. The polarization position angle map still displays the `spine-sheath' structure in the main jet, albeit with a reduced `spine', but in the counterjet the dominant orientation of the $E$-vector is now perpendicular to the axis. The `spine' shrinks because the emissivity rises toward the axis, so the largest contribution to the surface brightness is acquired where the line of sight passes closest to the axis, where the magnetic field projected on the plane of the sky is entirely parallel to the axis (tangential points in Figure \ref{pic_loop}). The naive conversion of the polarization position angle into the orientation of the magnetic field would imply that the field is mostly parallel to the axis, but in fact the fields are purely toroidal. This example illustrates that not only the orientation of the jet is important in inferring the direction of the magnetic field, but the emissivity profile as well.

In the core-dominated case, the polarization fraction is small in the center, where multiple field directions contribute, resulting in depolarization. The polarization fraction is close to its theoretical maximum on the edge, where only a thin layer of the jet material contributes to the observed emission. This rise is reminiscent of the puzzling polarization fraction profiles in the parsec-scale jets of 3C273 \citep{zava05} and in Mkn501 \citep{push05}. We suggest that in these objects the emissivity is rising toward the axis, producing the observed polarization pattern. In both these objects the orientation of the $E$-vector of polarization is perpendicular to the jet in the regions where the emission is highly polarized, in agreement with our predictions. To further strengthen the case for the toroidal fields in these objects, Mkn501 shows a well-defined `spine-sheath' polarization pattern, while 3C273 shows strong gradients of the rotation measure across the jet. 

The core-dominated case is only mildly relativistic (the highest Lorentz-factor on the edge of the jet is only $\gamma \sim 3$ in Figure \ref{pic_map}B), so our solutions for the lateral equilibrium are not, strictly speaking, very accurate in this case. However, the effects of the geometric projection and emissivity profile on emission and polarization maps are rather generic and are well-illustrated by this approximate model. 

\section{Conclusions}
\label{sec_conclusions}

We have considered axisymmetric relativistic MHD jets with an ultrarelativistic equation of state and toroidal magnetic fields. We have then found all self-similar solutions of relativistic MHD equations in this setup assuming that the jet propagates in a pressure equilibrium with the surrounding matter. There are three geometrically self-similar solutions which occur in the medium with pressure scaling as $P(z)\propto z^0$ (cylindrical jet), $\propto z^{-2}$ (parabolic jet) and $\propto z^{-4}$ (conical jet). The lateral structure of the jet is determined by the balance of the radial pressure gradient, the tension of the toroidal magnetic field lines and the centrifugal force. The last term is relevant only when the streamlines are curved. Our models represent a special case of a more general situation in which interactions between the jet and the medium lead to shocked layers \citep{brom07}; in our case, solutions with no shocks are possible. 

While the self-similarity condition is quite restrictive, in that only a limited set of geometries can be studied, the lateral structure of the parabolic jet displays some generic features which we investigate in detail. The pressure gradients act to push the jet material against the walls of the jet, whereas the magnetic field tension acts to pile the material closer to the axis. The relative importance of the two effects is determined by the physical parameters of the jet, such as its magnetization $\beta_B$ and the total energy. The pinched core component is stronger for a higher value of $\beta_B$, whereas the pile-up component is stronger for a jet with a higher Lorentz-factor (or higher total energy, to put it another way).  

Our models describe jets whose Lorentz-factors are smaller than the critical value $\gamma_{\rm cr}\simeq 1/\theta_j$, where $\theta_j$ is the opening angle of the jet. We find that for larger Lorentz-factors $\gamma\gg\gamma_{\rm cr}$ either the formal solution of the lateral equilibrium equation does not exist at all, or it exists in the form of a thin pile-up of the material along the edge of the jet. The condition $\gamma<\gamma_{\rm cr}$ is equivalent to requiring that the jet maintains causal contact in the lateral direction as it propagates.

Edge pile-up is a generic feature of a jet with curved streamlines and it might explain edge-brightening of the jet in M87 in the acceleration region \citep{ly07, kova07}. We also find that even purely toroidal magnetic fields can produce the `spine-sheath' polarization structure \citep{push05} when projection effects are taken into account. We find that for core-dominated jets the polarization fraction is expected to be near zero on the projected jet axis and rise toward the edge, as observed in 3C273 \citep{zava05} and Mkn501 (B1652+398; \citealt{push05}). The orientation of the polarization position angle is determined not just by the geometric projection effects, but also by the emissivity profile. For emissivity rising toward the axis, the polarization perpendicular to the axis may become the dominant orientation even for purely toroidal magnetic fields. These effects significantly complicate the interpretation of the observed polarization position angle in terms of the intrinsic direction of the magnetic field. 

\acknowledgments
The authors are very grateful to Chun Ly and collaborators for providing the map of radio intensity in M87 in electronic form (Figure 3 of \citealt{ly07}). NLZ would like to thank Anatoly Spitkovsky and Paul Wiita for discussions, the members at JILA (University of Colorado at Boulder) for hospitality, and Serguei Komissarov and Vasily Beskin for the comments on the manuscript.   

NLZ is supported by NASA Spitzer Space Telescope Fellowship Program, through a contract issued by the Jet Propulsion Laboratory, California Institute of Technology under a contract with NASA. MCB acknowledges support from NSF grant AST-0307502 and NASA Astrophysics Theory Program grant NNG06GI06G.

\begin{thebibliography}{}
\bibitem[Aharonian et al.(2005)]{ahar05}Aharonian, F., et al. 2005, A\& A, 442, 895
\bibitem[Aloy et al.(2000)]{aloy00}Aloy, M.-A., G\'omez, J.-L., Ib\'a\~nez, J.-M., Mart\'{\i}, J.-M., \& M\"uller, E. 2000, \apj, 528, L85
\bibitem[Anile(1989)]{anil89}Anile, A.M. 1989, Relativistic fluids and magneto-fluids (Cambridge; New York: Cambridge Univ. Press)
\bibitem[Asada et al.(2002)]{asad02}Asada, K., Inoue, M., Uchida, Yu., \& Kameno, S. 2002, PASJ, 54, L39
\bibitem[Attridge et al.(1999)]{attr99}Attridge, J.M., Roberts, D.H., \& Wardle, J.F.C. 1999, \apj, 518, L87
\bibitem[Begelman(1995)]{bege95}Begelman, M.C. 1995, Proc. Natl. Acad. Sci., 92, 11442
\bibitem[Begelman et al.(1984)]{bege84}Begelman, M.C., Blandford, R.D., \& Rees, M.J. 1984, Rev. Mod. Phys., 56, 255
\bibitem[Begelman \& Cioffi(1989)]{bege89}Begelman, M.C. \& Cioffi, D.F. 1989, \apj, 345, L21 
\bibitem[Begelman \& Li(1994)]{bege94}Begelman, M.C. \& Li, Z.-Y. 1994, \apj, 426, 269
\bibitem[Beskin \& Nokhrina(2006)]{besk06}Beskin, V.S. \& Nokhrina, E.E. 2006, \mnras, 367, 375
\bibitem[Blandford \& Payne(1982)]{blan82}Blandford, R.D. \& Payne, D.G. 1982, \mnras, 199, 883
\bibitem[Bromberg \& Levinson(2007)]{brom07}Bromberg, O. \& Levinson, A. 2007, \apj, accepted, arXiv:0705.2040
\bibitem[Contopoulos(1994)]{cont94}Contopoulos, J. 1994, \apj, 432, 508
\bibitem[Contopoulos(1995)]{cont95}Contopoulos, J. 1995, \apj, 450, 616
\bibitem[Dixon(1978)]{dixo78}Dixon, G. 1978, Special relativity, the foundation of microscopic physics (Cambrige; New York: Cambridge Univ. Press)
\bibitem[Dondi \& Ghisellini(1995)]{dond95}Dondi, L. \& Ghisellini, G. 1995, \mnras, \bibitem[Gabuzda et al.(2004)]{gabu04}Gabuzda, D.C., Murray, E., \& Cronin, P. 2004, \mnras, 351, L89
\bibitem[Gabuzda et al.(2000)]{gabu00}Gabuzda, D.C., Pushkarev, A.B., \& Cawthorne, T.V. 2000, \mnras, 319, 1109
\bibitem[Gammie et al.(2003)]{gamm03}Gammie, C.F., McKinney, J.C., \& T\'{o}th, G. 2003, \apj, 598, 444
\bibitem[Ginzburg(1979)]{ginz79}Ginzburg, V.L. 1979, Theoretical Physics and Astrophysics (Oxford; New York: Pergamon Press)
\bibitem[Hardcastle et al.(2002)]{hard02}Hardcastle, M.J., Worrall, D.M., Birkinshaw, M., Laing, R.A., \& Bridle, A.H. 2002, \mnras, 334, 182
\bibitem[Horiuchi et al.(2006)]{hori06}Horiuchi, S., Meier, D.L., Preston, R.A., \& Tingay, S.J. 2006, PASJ, 58, 211
\bibitem[Jorstad et al.(2005)]{jors05}Jorstad, S.G., et al. 2005, \aj, 130, 1418
\bibitem[Junor et al.(1999)]{juno99}Junor, W., Biretta, J.A., \& Livio, M. 1999, Nature, 401, 891
\bibitem[Komissarov(1999)]{komi99}Komissarov, S.S. 1999, \mnras, 308, 1069
\bibitem[Komissarov et al.(2007)]{komi07}Komissarov, S.S., Barkov, M.V., Vlahakis, N., \& Konigl, A. 2007, \mnras, 380, 51
\bibitem[Kovalev et al.(2007)]{kova07}Kovalev, Y.Y., Lister, M.L., Homan, D.C., \& Kellermann, K.I. 2007, \apjl, in press; arXiv:0708.2695
\bibitem[Laing \& Bridle(2002)]{lain02}Laing, R.A. \& Bridle, A.H. 2002, \mnras, 336, 1161
\bibitem[Li et al.(1992)]{li92}Li, Z.-Y., Chiueh, T., \& Begelman, M.C. 1992, \apj, 394, 459
\bibitem[Lister \& Homan(2005)]{list05}Lister, M.L. \& Homan, D.C. 2005, \aj, 130, 1389
\bibitem[Ly et al.(2007)]{ly07}Ly, C., Walker, R.C., \& Junor, W. 2007, \apj, 660, 200
\bibitem[Lyutikov et al.(2003)]{lyut03}Lyutikov, M., Pariev, V.I., \& Blandford, R.D. 2003, \apj, 597, 998
\bibitem[Lyutikov et al.(2005)]{lyut05}Lyutikov, M., Pariev, V.I., \& Gabuzda, D.C. 2005, \mnras, 360, 869
\bibitem[Mirabel \& Rodr\'{\i}guez(1999)]{mira99}Mirabel, I.F. \& Rodr\'{\i}guez, L.F. 1999, \araa, 37, 409
\bibitem[Pariev et al.(2003)]{pari03}Pariev, V.I., Istomin, Ya.N., \& Bereznyak, A.R. 2003, A\& A, 403, 805
\bibitem[Perucho et al.(2007)]{peru07}Perucho, M., Hanasz, M., Mart\'{\i}, J.-M., \& Miralles, J.-A. 2007, Phys. Rev. E, 75, 056312
\bibitem[Pushkarev et al.(2005)]{push05}Pushkarev, A.B., Gabuzda, D.C., Vetukhnovskaya, Yu.N., \& Yakimov, V.E. 2005, \mnras, 356, 859
\bibitem[Piran(1999)]{pira99}Piran, T. 1999, Phys. Rep., 314, 575
\bibitem[Stone \& Hardee(2000)]{ston00}Stone, J.M. \& Hardee, P.E. 2000, \apj, 540, 192
\bibitem[Ustyugova et al.(1995)]{usty95}Ustyugova, G.V., Koldoba, A.V., Romanova, M.M., Chechetkin, V.M., \& Lovelace, R.V.E. 1995, \apj, 439, L39
\bibitem[Ustyugova et al.(2000)]{usty00}Ustyugova, G.V., Lovelace, R.V.E., Romanova, M.M., Li, H., \& Colgate, S.A. 2000, \apj, 541, L21
\bibitem[Vlahakis \& K\"onigl(2003a)]{vlah03a}Vlahakis, N. \& K\"onigl, A. 2003a, \apj, 596, 1080
\bibitem[Vlahakis \& K\"onigl(2003b)]{vlah03b}Vlahakis, N. \& K\"onigl, A. 2003b, \apj, 596, 1104
\bibitem[Zavala \& Taylor(2005)]{zava05}Zavala, R.T., \& Taylor, G.B. 2005, \apj, 626, L73
\end{thebibliography}
\end{document}